\documentclass[usenatbib,useAMS]{mn2e} \usepackage{times}
\usepackage{graphicx}
\usepackage{natbib}

\title[Modelling galactic spectra: II - Simultaneous study of stellar
dynamics and stellar mix in NGC~3258]{Modelling galactic spectra: II -
Simultaneous study of stellar dynamics and stellar mix in
NGC~3258\thanks{Based on observations obtained at the European Southtern Observatory, La Silla, Chile (Programmes Nr. 62.N-0492, 64.N-0192)}}
\author[V. De Bruyne et al.]
       {V. De Bruyne$^{1}$, S. De Rijcke$^{1}$\thanks{Postdoctoral Fellow of the Fund for Scientific
Research - Flanders (Belgium)(F.W.O)}, H. Dejonghe$^{1}$\thanks{E-mail:Herwig.Dejonghe@rug.ac.be}, W.W. Zeilinger$^{2}$ \\
        $^{1}$Astronomical Observatory, Ghent University, Krijgslaan 281, S9, 9000 Ghent, Belgium\\
$^{2}$ Institut f\"ur Astronomie, Universit\"at Wien, T\"urkenschanzstrasse 17, A-1180 Wien, Austria}

\date{Accepted 
      Received ;
      in original form }

\pagerange{\pageref{firstpage}--\pageref{lastpage}}
\pubyear{}

\begin{document}

\maketitle

\label{firstpage}

\begin{abstract}

In this paper we adopt a method to analyse absorption line spectra
from elliptical galaxies that determines the dynamics of the galaxy
and at the same time offers a way to study the stellar populations in
that galaxy by a direct fit to the spectra.  The result of the
modelling is a distribution function for each stellar population that
is considered in the fit. The method is described in detail in an
accompanying paper \citep{pap1}.

This paper reports on a dynamical stellar population study in
NGC~3258, based on spectra in two different wavelength regions, the
near-IR Ca II triplet around 8600 {\AA} and the Ca H and K lines around
3900 {\AA}.  These absorption lines have discriminating power toward
various stellar types. 

The dynamical model shows an increase in dwarfs (represented by G2V
stars) toward the centre. Most of the rotation in the model is
delivered by the giants (represented by M1III stars). Moreover, the
different models that were considered indicate that establishing a
potential for a galaxy is dependent on the wavelength range used
for the modelling.

\end{abstract}


\section{Introduction} \label{intro}

Elliptical galaxies are no longer seen as simple isothermal objects
that follow a de Vaucouleurs-law and that are chemically well mixed.
Color gradients indicating that elliptical galaxy centres have redder
colours than the outer regions (e.g. \citet{mi}, \citet{pel1},
\citet{pel2}), and variations in line strengths (see \citet{ka} and
references therein) can be interpreted in terms of a metallicity or an
age gradient \citep{wo}.  Moreover, spectroscopic studies revealed
that a considerable number of early-type galaxies have a kinematically
distinct core, that may have a separate stellar population
(\citet{bert}, \citet{sur}, \citet{meh}, \citet{drdc}).

Current stellar population modelling techniques for elliptical
galaxies explicitly take an evolution scenario into account.
Most of them use theoretically calculated isochrones and a library of
theoretical or observed stellar spectra.
A number of frequently used techniques are:
\begin{itemize}
\item Evolutionary synthesis models (\cite{brch}, \citet{wo}, \citet{br}) :
these models predict the spectral evolution of a stellar population,
based on computed isochrones (i.e. the positions of the stars of a
single-age, single-metallicity population in a color-magnitude diagram
at a given instant). In such a single-burst population, the
distribution of the stars over mass is given by the initial-mass
function (IMF). The spectral properties of a population in which stars
have formed over an extended period of time according to some
star-formation rate, can be calculated as convolution integrals of the
corresponding properties of the single-burst populations. Hence the
need for a large library of observed or computed spectra of stars of
various ages, masses and metallicities.

\item Chemo-evolutionary models (\citet{br}, \citet{vaz1},
\citet{vaz2}) : these models are similar to the previous ones except
that a physical basis is provided for the star-formation rate. The
evolution of a galaxy is followed starting from a uniform gas cloud in
which stars form at a rate given by the Schmidt law which assumes the
star-formation rate to be proportional to the gas fraction. Both the
evolutionary synthesis models and the chemo-evolutionary models assume
elliptical galaxies to be homogeneous at all times so that they cannot
describe metallicity gradients.

\item Chemo-dynamical models (\citet{ber}, \citet{the}) : these
sophisticated models use a Smoothed Particle Hydrodynamics (SPH) code
to describe the dynamics of the gaseous phase in a galaxy. Within this
framework, star formation, energy feedback and chemical enrichment can
be incorporated, providing a realistic description of the chemical and
dynamical evolution of galaxies. As the assumption of homogeneity is
relaxed, metallicity gradients can be studied.
\end{itemize}



The models mentioned above produce colors and absorption line indices.
Despite the fact that it is hard to overcome the age-metallicity
degeneracy when disentangling the integrated galaxy light into
separate stellar populations, it seems that elliptical galaxies can be
well represented by a single-age, single-metallicity model
(e.g. \citet{brch}, \citet{wo}, \citet{vaz1}). This suggests that most
of the stars in these early-type galaxies were formed in a single
burst, long ago. More recently \citet{trag1} and \citet{trag2}
presented data and models for the centers of elliptical galaxies and
found a wide range of stellar population parameters. Moreover, their
best fitting models implied adding a younger stellar population to an
underlying old population.

On the other hand, the calculation of line widths, needed for
comparison between observations and population synthesis models,
suffers from the fact that the internal velocity dispersion results in
a broadening of the spectral lines. This affects both the continuum
and the lines, and at high velocity dispersion cause the
absorption-line indices to appear weaker than they really are. This
results in a decrease of the measured equivalent widths.  Therefore,
some sort of velocity dispersion correction is applied to the
equivalent widths calculated from the synthesis models. The velocity
dispersion used for this correction is in most cases taken to be the
$\sigma$ from the Gaussian that is used as model line-of-sight
velocity distribution (hereafter LOSVD) (\citet{vaz3}, \citet{wo}),
which is now believed to be only a first approximation of the true
LOSVD. Furthermore, metallicity gradients in ellipticals are in most
cases determined without taking the radial gradient in the velocity
dispersion into account \citep{vaz3}, or by using some simple scaling
relation.

In this paper the possibility to exploit the tight connection between
stellar populations and stellar dynamics is explored. Dynamics and
stellar populations are studied by modelling directly the observed
spectra \citep{dr}. The dynamical model holds a distribution function
(hereafter DF) that gives the densities of the stars in phase
space. Fitting directly to the galactic spectra has the consequence
that a DF is known for each stellar template that is considered in the
fit. That such a method is potentially rewarding, has been proved
by the seminal work by Pickles (\citet{pivi}) and by the
extensive tests we have performed using synthetic data. These DFs are
sums of basis functions and are obtained by a quadratic programming
code that minimizes the differences between the observed spectra and
the spectra calculated from the model, taking positivity constraints
for the DF into account.

Whereas a first paper in this series \citep{pap1} (hereafter paper I)
elaborated on the use of this method for establishing a gravitational
potential for the elliptical galaxy NGC~3258, based on spectra of the
near-IR CA II triplet, this paper has the scope to investigate the
behaviour of different stellar classes in this galaxy. For this
reason, the modelling uses spectral features from different wavelength
ranges, the near-IR Ca II triplet and the blue Ca H and K lines, that
are believed to be tracers of different stellar populations.  For an
extensive discussion of the modelling strategy and the basis functions
that were used to construct the DFs, the reader is referred to paper
I.

The observations of the Ca H and K lines near 3900 {\AA} are presented
in section 2. These data will be referred to as the 'B' data, in
contrast with the data set around 8600 {\AA}, which will be called the
'near-IR' data. The data reduction and model for this data set is
presented in paper I. In section 3, we discuss two of the tests
we have performed to show the method's potential. Section 4 gives an
overview of the absorption features and stellar templates that were
used. As a first step, the B data set is modeled separately. The
results for the potential and the stellar templates can be compared
with the results in paper I, this is done in section 5. The model for
the combined data set is presented in section 6. A discussion and
conclusions can be found in section 7 and 8 respectively.

\section{Observations and data reduction}

The E1 galaxy NGC~3258 is found at a distance of 36.12 Mpc (taking
$H_0$=75 km/s/Mpc), hence 1 kpc corresponds to $5.7''$.  The galaxy is
a member of the Antlia Group and has NGC~3260 (at $2.6'$) and
NGC~3257 (at $4.5'$) in its neighbourhood.

The photometric data are presented in paper I. There is also a dust
disk.

\subsection{Spectroscopy}
Longslit spectra along the optical major axis ofNGC~3258 were obtained
with the ESO-NTT telescope in the nights of 11-12/2/1999, using the
blue arm of EMMI, covering the Ca H and K lines around 3900 {\AA}.  A
number of standard stars (G dwarfs and K and M giants) were also
observed in the same instrumental setup, they are listed in table
\ref{temps}.

For these spectra, grating \# 12 was used, having a dispersion of 0.92
{\AA}/pix. The detector was a Tektronix CCD with 1024$\times$1024
pixels, 24$\mu m \times$24$\mu m$ in size and with a pixel scale of
$0.37''$/pix. A slit width of $1.5''$ yielded a spectral resolution of
3.73 {\AA} FWHM, resulting in an instrumental broadening of about 122
km/s in the region of the Ca H and K lines. Several exposures of 3600 sec
were taken (in total 8 hours).

\begin{table}
\begin{tabular}{l|l}
\hline
HD 21019 & G2V\\
HD 115617&G5V\\
HD 102070 & G8III\\
HD 117818&K0III\\
HD 114038&K1III\\
HD 44951 & K3III\\
HD 29065&K4III\\
HD 99167 & K5III\\
HD 129902&M1III\\
HD 93655&M2III\\
\hline
\end{tabular}
\caption{Observed stellar templates.}
\label{temps}
\end{table}

Standard reduction steps were
applied to these spectra with ESO-MIDAS\footnote{ESO-MIDAS is
developed and maintained by the European Southern Observatory}.  A
more detailed description of the data reduction can be found in paper
I. The calibrated spectra containing the Ca H and K lines have a step of
$0.45$ {\AA}.



In a second observation run (in the nights of 27-28/2/2000) major axis
spectra taken using the red arm of EMMI, covering the Ca II triplet
around 8600 {\AA} (see paper I).  Care was taken to put the slit on
the same position of the galaxy as in the previous observing run in
order to be able to compare the data sets. The galaxy center, being
the brightest spot of the galaxy, was taken as reference. The accuracy
was better than the equivalent size of a pixel.

\subsection[]{Kinematic parameters}\label{data}

\begin{figure}
\includegraphics[bb=30 150 520 560,clip=,scale=.4]{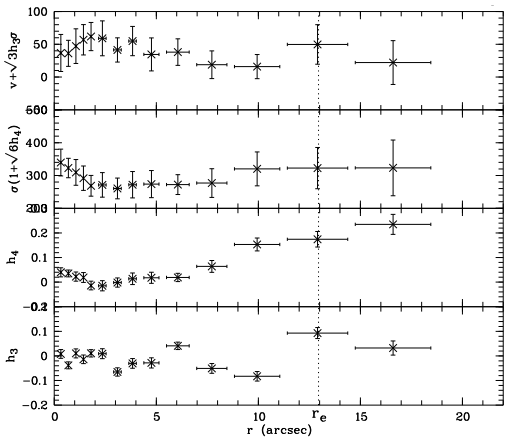} \caption{Kinematics obtained from
  the spectral feature around 3900 {\AA}.}
\label{kin}
\end{figure}

In figure \ref{kin} kinematic parameters from the B data set are
presented.  These data are retrieved from the observations in the same
way as the near-IR data. For a more detailed description of the
latter, the reader is referred to paper I. The spectra for the B data
set are analysed with a mix of a G5V dwarf and a M1III giant.

It is not surprising that the values for the kinematic parameters are
not exactly the same for the blue data and the near-IR data.  There
can be various reasons for the this, like differences in data
reduction, the fact that different stellar populations are looked at
or that for none of the data set a perfectly matching template was
used.  Such differences are also reported in \cite{bar}. These authors
use Mg and Fe lines to derive velocity dispersions. They argue that
the results for the blue dispersions may suffer from the fact that
elliptical galaxies have a different Mg/Fe ratio than the stellar
templates that are used and that these results cause the difference
between the blue and red velocity dispersions. They suggest it would
be better to only fit the Mg or Fe lines. For the velocity dispersions
presented in this paper Ca lines were used in both wavelength
regions. Hence, in contrast with \cite{bar} the risk of using template
stars with different abundance ratios is not existing in this case.
Although there is a slight systematic offset, both kinematic parameter
sets are in agreement, in the sense that the data mostly lie within
each others error bars.


\section{Testing the modeling method}

As was argued in section \ref{intro}, the composition of the stellar
population of elliptical galaxies exhibits radial gradients. Hence the
introduction of multiple distribution functions : these allow to model
a galaxy with a radially varying stellar mix in which stars with
different physical characteristics can be in different dynamical
states. Of course, it will only be possible to discriminate between
stars that have sufficiently different spectral features in the
wavelength region(s) covered by the data. Therefore, the use of a
broad spectral range, containing absorption lines that are strong
enough to show up in noisy galaxy spectra as well as being
sufficiently sensitive to the composition of the stellar mix will be
prerequisite. Before applying it to real data, we have extensively
tested the modeling method and the robustness of the results using
synthetic spectra covering a 200~{\AA}-wide region around the Mg{\sc
i} triplet and in two regions (8478\AA-8570\AA~and 8638\AA-8700\AA)
containing the lines of the Ca{\sc ii} triplet. We will disscuss the
most relevant results of two of these tests.

\begin{figure}
\vspace{5cm}
\special{ hscale=30 vscale=30 hsize=570 vsize=150 
	 hoffset=30 voffset=-5 angle=0 psfile="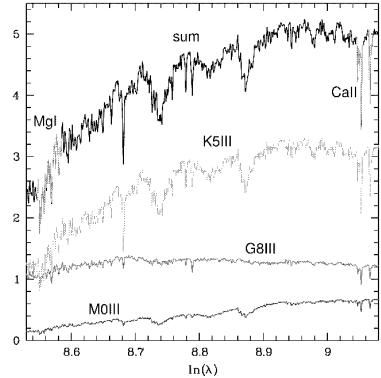"}
\caption{Test 1 : the contributions of the G8{\sc iii} class stars,
the K5{\sc iii} stars and the M0{\sc iii} stars to the central line of
sight of the synthetic spectrum. The sum of these contributions is
plotted in black. Only the regions around the Mg{\sc i} and Ca{\sc ii}
lines were used. \label{popf1}}
\end{figure} 
\begin{figure*}
\vspace{9cm}
\special{ hscale=50 vscale=50 hsize=520 vsize=260 
	 hoffset=10 voffset=285 angle=-90 psfile="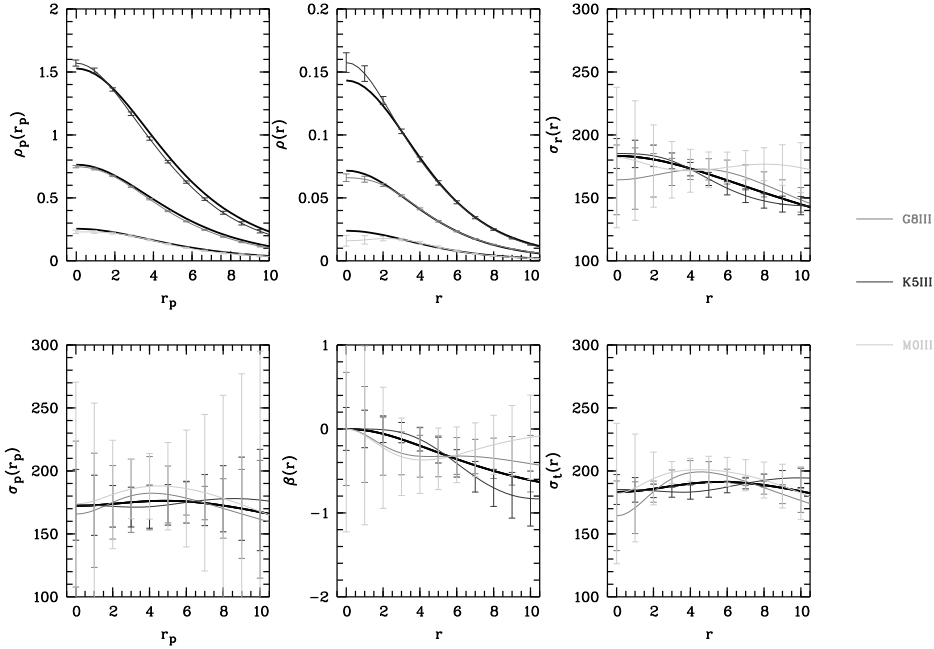"}
\caption{{\bf Test 1}:~the kinematics of the three stellar classes are
compared to the input Plummer model (the surface brightness, $\rho_p$,
the projected velocity dispersion $\sigma_p$, the spatial density,
$\rho$, the radial and tangential velocity dispersions, $\sigma_r$ and
$\sigma_t$ and Binney's anisotropy parameter $\beta$). Density units
are arbitrary, velocity dispersions are expressed in km/s, linear
distances in kpc and projected distances in arcseconds. The kinematics
of the stellar populations of the input model are plotted in bold
lines. The kinematics of the stellar populations of the fitted model
are over-plotted grey, as indicated in the figure. The kinematics of
the most luminous classes are best constrained : the errors on the
radial and tangential velocity dispersions of the G8{\sc iii} and the
K5{\sc iii} class are nowhere larger than 10-20~km/s whereas those of
the M0{\sc iii} class can be as large as 30~km/s. The projected
velocity dispersion of the three classes is retrieved with an accuracy
of about 10-20~km/s.
\label{popf1kin}}
\end{figure*} 

\begin{figure*}
\vspace{9cm}
\special{ hscale=50 vscale=50 hsize=520 vsize=260 
	 hoffset=10 voffset=285 angle=-90 psfile="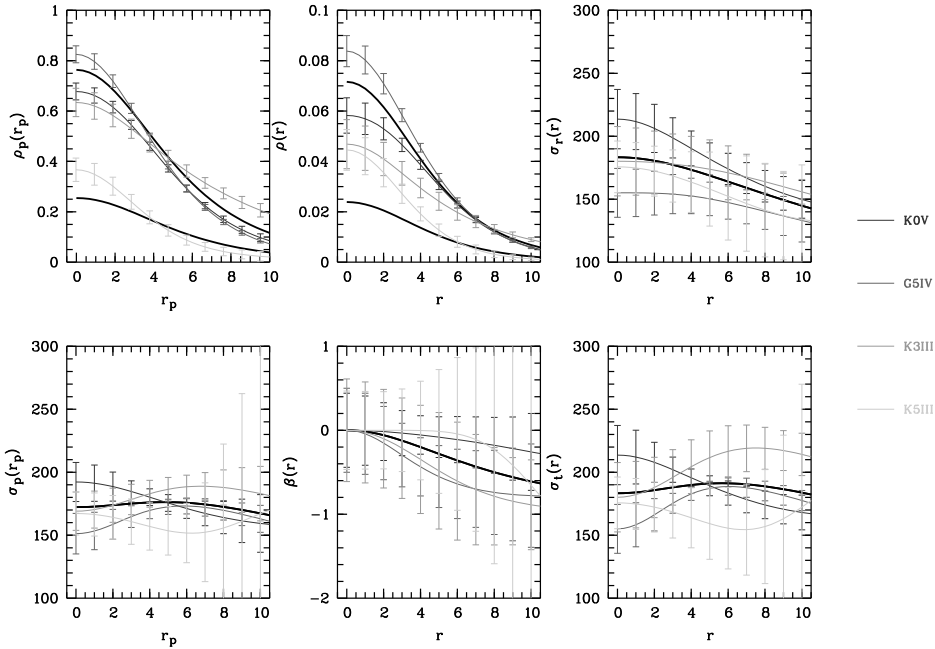"}
\caption{{\bf Test 2}:~the kinematics of the input model are plotted
in bold lines for the G5{\sc iv}, K0{\sc v}, K3{\sc iii}, and M0{\sc
iii} class. The kinematics of the fitted model are over-plotted in
grey for the G5{\sc iv}, K0{\sc v}, K3{\sc iii}, and K5{\sc iii}
class. The latter replaces the M0{\sc iii} class of the input model to
mimic template mismatch. The same quantities are presented as in
Fig. \ref{popf1kin}, using the same units. The deviations between the
fitted model and the input model are clearly much larger than those in
Fig. \ref{popf1kin}. The template mismatch results in the fitted model
having a significantly too high a $\chi^2$-value, allowing the model
to be rejected at the 99\% level. \label{popf5kin}}
\end{figure*} 

As a first test, the modeling method is applied to synthetic spectra
of a tangentially anisotropic Plummer model. The spectrum of this
model galaxy is calculated using the spectra of a G8{\sc iii}, a
K5{\sc iii} and a M0{\sc iii} star (see Fig. \ref{popf1}). These
stellar spectra cover a very wide spectral range (4800\AA-8920\AA)
with high resolution (1.25\AA). They are taken from the AAS CDROM,
volume {\sc vii}, 1996, \cite{leith}. The synthetic spectrum extends
to one half-light radius of the Plummer model \citep{dj}. The signal-to-noise
ratio is about 130 per pixel in the central line-of-sight (a pixel
measures 0.5$''$ in the spatial direction and 50~km/s in the
dispersion direction), dropping to 60 per pixel at the last
data-point. The stellar spectra that are used to calculate the
synthetic data also serve as template spectra in the modeling. Thus,
we need not worry about template mismatch. We also adopted the correct
gravitational potential to model the data. Only the constraint that
the distribution function of each stellar class must be positive is
imposed.

The results of this experiment are presented in figure \ref{popf1kin},
where we compare the retrieved kinematics of the three stellar classes
with the input model. The projected and the spatial densities are
reproduced rather well. The kinematics of the G8{\sc iii} and K5{\sc
iii} classes are retrieved with comparable accuracy. The deviations
between the radial and tangential velocity dispersions of the fitted
model and the input model are everywhere between 10-20~km/s. The
corresponding errors of the M0{\sc iii} class can be as large as
30~km/s. The projected dispersions of the three classes can be
reproduced with an accuracy of about 10-20~km/s. The errorbars that
express the uncertainties, introduced by the noise on the data, on the
kinematics of the separate classes are large. This is quite natural
since only the summed properties of the three classes are constrained
by the data. The errorbars on the kinematics of the total stellar
population are much smaller. Concluding, this experiment yields a
rather satisfactory result. Given ideal circumstances -- the correct
potential is assumed, template mismatch is no worry -- the modeling
technique is able to give a fairly reliable idea of the dynamical
state of the three stellar classes in the model galaxy.

As a second test, we use a more varied stellar mix and we do not use
the stellar spectra that are used to calculate the synthetic data as
templates. The synthetic spectra are calculated with K0{\sc v}, G5{\sc
iv}, K3{\sc iii} and the M0{\sc iii} stellar spectra while we use a
K5{\sc iii} instead of a M0{\sc iii} template spectrum when fitting to
the spectra. The other three template spectra are left untouched. A
number of low-order continuum terms are added to the library of
components to help the modeling method compensate for this template
mismatch. In the spectral region around the Mg{\sc i} triplet, the
K5{\sc iii} and the M0{\sc iii} are almost indistinguishable. The
lines of the Ca{\sc ii} triplet are about 10\% deeper in the M0{\sc
iii} than in the K5{\sc iii} spectrum. In both spectra, these lines
are equally broad.

The results of this test are presented in figure \ref{popf5kin}. The
K5{\sc iii} template more or less takes over the role of the M0{\sc
iii} spectrum. Since the absorption lines in the K5{\sc iii} and the
M0{\sc iii} template spectrum are equally broad, no large systematic
errors are introduced in the kinematics : the spatial velocity
dispersions are reproduced with an accuracy of 30-40~km/s. The spatial
densities of the different stellar classes are only poorly
retrieved. This is a corollary of the fact that the model needs to
compensate for the differences in the continuum shape and the depth of
the absorption lines between the K5{\sc iii} and the M0{\sc iii}
spectrum. An important conclusion of this test is that we pay dearly
the systematic deviations between the galaxy spectrum and the model
spectrum in the Ca{\sc ii} triplet region in terms of the
$\chi^2$. Its value is significantly too high, allowing the model to
be rejected at the 99\% level. Thus, the modeling method is sensitive
to the composition of the stellar population~: one is able to reject a
galaxy model if the spectral properties of its stellar population do
not match those of the galaxy's stellar mix. Again, the importance of
a judicious choice of the spectral regions included in the fit is
demonstrated. If only the Mg{\sc i} triplet region is used in the fit,
the model provides an excellent fit to the data, reflected in a
perfectly acceptable $\chi^2$ value.

As a conclusion, one can state that it is possible to extract
information from galaxy spectra on the dynamics of different stellar
classes if the modeling method is applied to spectral regions that
contain absorption features that are sensitive to the composition of
the stellar population. If the experiments described in this paragraph
are generic, a model can be expected to give a realistic description
of the composition and the dynamics of the stellar population of a
galaxy in the region where the data quality is has a $S/N$ of more
than 30 per bin.

\section{Spectral features and templates used for the modelling}
\subsection{Spectral features}
For the extraction of kinematic data out of galaxy spectra, one uses
absorption lines that are strong enough to show up in noisy galaxy
spectra. Most kinematic data available in the literature come from
features in the optical. In this region, the contribution of the sky
lines to the integrated spectra is moderate. Near the near-IR-region,
the sky becomes more prominent.  On the other hand, a stellar
population synthesis benefits from the use of a broad spectral region,
covering many absorption features.

For the simultaneous study of kinematics and populations as presented
in this paper it is important to use absorption features that have
enough signal and that have discriminating power toward various
stellar types. On the other hand, the number of data points that can
be included in the fit is limited.

For our modelling, we used the two strongest lines from the near-IR Ca
II triplet. These absorption features
are strong in the spectra of giant stars (F-M).

\citet{cen} find a complex behaviour of the Ca II triplet strength as
a function of three atmospheric parameters: effective temperature,
gravity and metallicity. For hot and cold stars, these lines are
rather insensitive to metallicity and can be used as a dwarf/giant
discriminant. This and the fact that they are very strong and
isolated, makes the lines very suited for kinematic studies
(\citet{dh}, \citet{km}).  Recent results indicate that these lines
are insensitive to age for populations older than 1 Gyr.  However,
they can be contaminated by TiO bands. These bands are absent in the
spectra of stars of spectral classes O till late K but become very
strong in spectra of very cool stars (later than M0).  

From the B data set, the Ca H and K lines were used. These two lines are
very strong in early type stars and are sensitive to
metallicity. These lines can be used as tracers for recent
starbursts (\citet{long1}, \citet{long2}).

The absorption lines used in the modelling are sensitive to different
stellar properties and can be used as a diagnostic for different
stellar populations. They give complementary information since they
are not influenced in the same way by the presence of dust in the
galaxy and their detection suffers in a different way from the night
sky.

Tests with a broader basis in wavelength, also including
the Ca4227, the G band and the Fe4383 feature, are presented and
commented on in \ref{techn}.

\subsection{Templates}
\begin{figure}
\includegraphics[scale=.4]{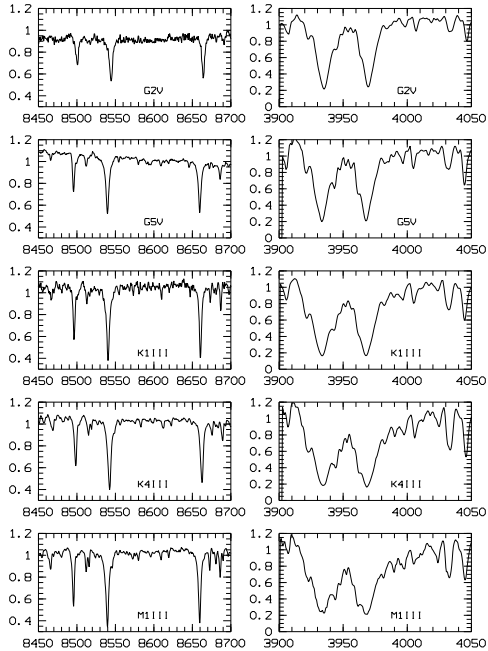}
  \caption{Star spectra used as templates around the Ca II triplet (left) and the Ca H and K lines (right), normalized and plotted on the same scale.}
 \label{temp}
\end{figure}

Five template stars that have discernible spectral properties were
used: a G2V dwarf, a G5V dwarf, a K1III giant, a K4III giant and a
M1III giant (listed in table \ref{temps}). The spectra are shown in
figure \ref{temp}.

The dynamical models were created in an iterative process.  First,
models were calculated with a single template star, using a library of
about 100 dynamical components.  Afterward, the components that are
selected for the single template models are used to create mixed
stellar template dynamical component libraries. With these mixed
libraries, new models were calculated.

\section[]{Models based on Ca H and K lines around 3900 {\AA}} 
The dynamical modelling strategy and the dynamical components that
were used, are discussed in paper I. In that paper, the models for
NGC~3258 based on the two strongest near-IR CaII triplet lines (at
8542 {\AA} and 8662 {\AA}) were presented. In similar fashion, the
gravitational potential can also be established from modelling the Ca
H and K lines.

\subsection{Potential}
The luminous mass density was obtained from the R luminosity density,
hence it is the same profile as used for modelling the Ca II triplet
features in paper I.

On a grid of models with different slope of dark matter halo and
different total mass, the lowest $\chi^2$ was obtained for a potential
where the spatial density of the total amount of matter was 2.5 times
the spatial density of the luminous matter at 2 kpc. This scaling
implies that the total mass at  $1 r_e$ is $1.3\times10^{11} M_\odot$,
$70 \%$ of which is luminous matter. The total mass at  $2 r_e$ is
$4.1\times10^{11} M_\odot$, and $37 \%$ of this is luminous matter.

Just like in paper I it was not necessary to include a black hole in
the modelling.  The model based on the near-IR data (paper I) used a
potential with the same relative amount of luminous and dark matter
but with a mass at $1r_e$ of $1.6\times 10 ^{11} M_\odot$.

\subsection{Dynamical model}
\begin{figure*}
\includegraphics[scale=.8]{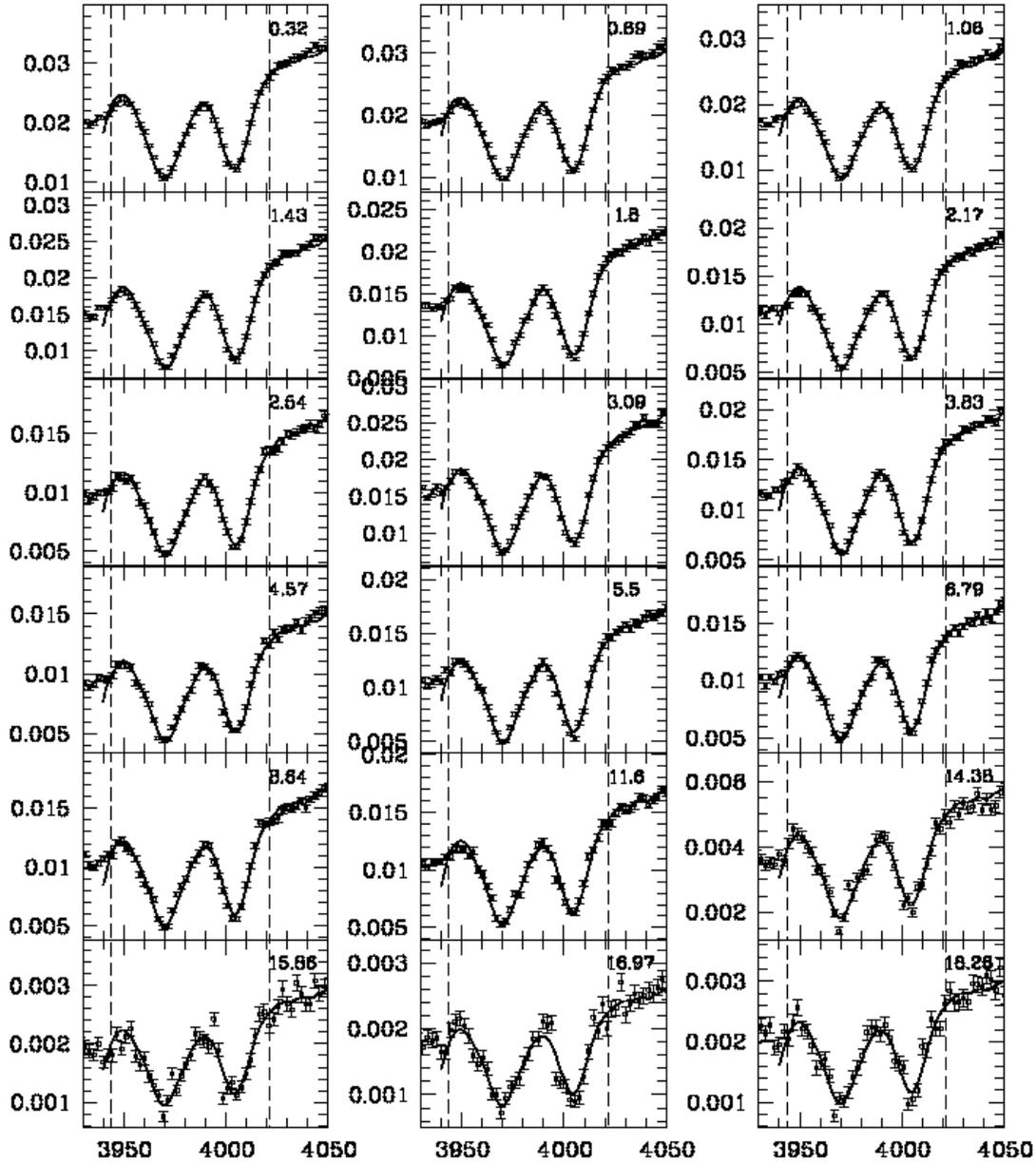} \caption{Data and model
  around the Ca H and K lines (B data set). The region included in the
  fit is located between the interval indicated by dashed lines. The
  projected radii of the positions where the spectra are taken are
  indicated in the panels.}  \label{specblue}
\end{figure*}

The best fitting model for the B data around the Ca H and K lines is
shown in figure \ref{specblue}. Judging from these plots, the fit
reproduces the spectra well. The error bars in the centre are very
small compared to the error bars in figure 12 in paper I.  The main
reason for this is that the Ca H and K lines are more pronounced
features in the spectrum than the Ca II triplet lines. Moreover, also
the contribution of the sky is much smaller in this region.
 
\begin{figure}
\includegraphics[bb=40 430 460 690,scale=.6,clip=true]{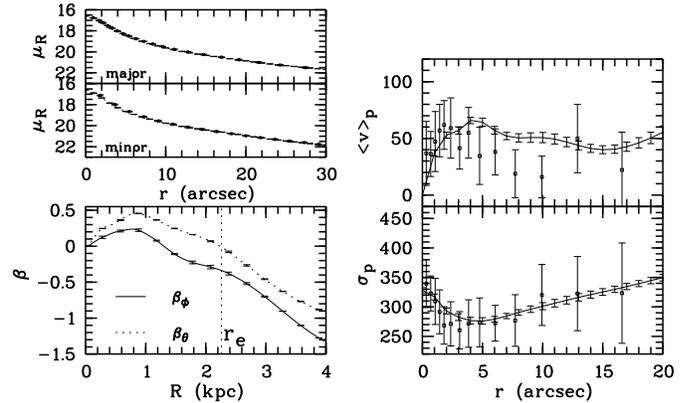}
 \caption{Some derived moments from the B model. Left column: projected
 density on major and minor axis and the anisotropy parameters
 $\beta_\phi$ and $\beta_\theta$. Right column: projected mean velocity
 and projected velocity dispersion.}  \label{modblue}
\end{figure}

Figure \ref{modblue} shows some derived quantities of the B model. The
upper left panel shows the projected density on major and minor
axis. As can be seen, the flattening of the galaxy is well reproduced.
The right panels show calculated projected mean velocity and velocity
dispersion profiles (solid lines), together with uncertainties on
these quantities as derived from the B model. Values and error
estimates for the corrected projected mean velocity and velocity
dispersion, as presented in figure \ref{kin} are plotted on top in
dots.

The mean velocity from the B model is between 5 and 12 arcsec
somewhat higher than the data points suggest.
In figure \ref{modblue} also the anisotropy parameters $\beta_\phi =
1-(\sigma^2_\phi/\sigma^2_r)$ (in solid line, also known as Binney's
anisotropy parameter) and $\beta_\theta =
1-(\sigma^2_\theta/\sigma^2_r)$ (in dotted line) are displayed in the
lower left panel.  For this model, $\beta_\phi$ is positive out to
about 1.2 kpc and drops to -1.3 at 4 kpc. This means that the model
changes from isotropic in the very centre to radial anisotropic and
further on to tangential anisotropic.

\begin{figure}
\includegraphics[bb=50 65 250 160,scale=.8,clip=true]{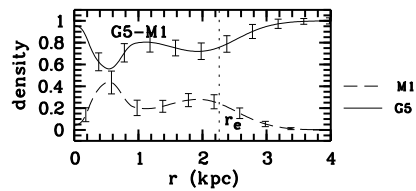} 
\caption{Template mix of the B model composed of G5V and M1III stars.}
\label{mixblue}
\end{figure}

Figure \ref{mixblue} shows the resulting template mix for the best
fitting model. In this case the combination of G5V and M1III stars
yielded the best result. At all positions, the largest fraction of the
density can be found in the G5V stars. The fraction of M1III stars is
more than 10 $\%$ between 0.2 and 2.8 kpc, with a peak of 45 $\%$ near
0.5 kpc.  Further tests have shown that a model with G5V and K4III
stars clearly preferred G5V stars, only in the centre there is a very
slight contribution of K4III stars. Models where G2V stars are used
instead of G5V stars yield the same conclusions concerning mixing of
dwarfs and giants, although the relative densities are different.

\section[]{Models based on both spectral regions.}\label{both}

\subsection{Potential}
The best model for the Ca II triplet data and the best model for the
Ca H and K data, were based on a potential with an equal relative amount
of dark matter, but the total masses for the models were different.
Therefore, also the model based on the combined data sets will be
modeled with a potential that is scaled such that the total amount of
matter is 2.5 times the amount of luminous matter at 2 kpc.  The
near-IR model based on the Ca II triplet feature required a total mass
that is 20 \% higher than that of the model based on the Ca H and K
lines.  For the model based on the combined spectra, a potential with
intermediate mass profile was used.

\subsection{The fit}
\begin{figure*}
\includegraphics[scale=.8]{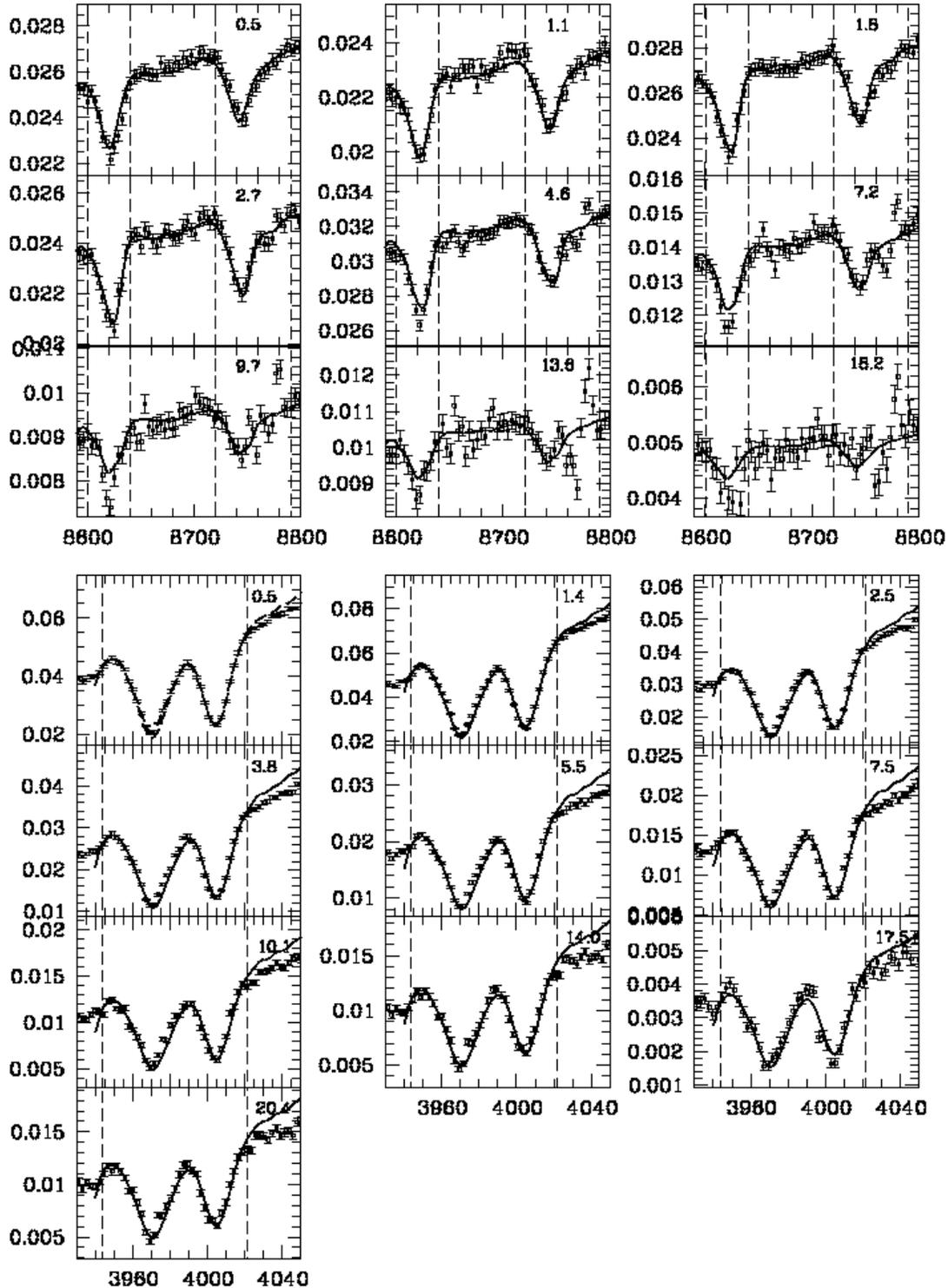} \caption{Data and model for
  both spectral regions: around the Ca II triplet lines in the first 9
  panels, around the Ca H and K lines in the next 10 panels. The parts
  included in the fit are between dashed lines. The projected radius
  of the spectra is indicated in the panels.}  \label{specbr}
\end{figure*}

Figure \ref{specbr} shows the simultaneous fit to the spectra around
8600 {\AA} (first 9 panels) and around 3900 {\AA} (last 10
panels). The spectra were rebinned to higher S/N than for the near-IR
and B models based on one spectral region, in order to keep the number
of data points in the fit about the same. Overall, the spectral
features are well represented.

The minor differences in quality between the fit of this model and
each of the models based on only one data set is probably due to the
fact that the total mass for this model was different from the total
masses used for the near-IR and B models based on only one spectral
region. The total mass is 8.3 $\%$ lower than the mass for the near-IR
model based on the 8600 {\AA} data and 10 $\%$ higher than the mass
for the B model based on the 3900 {\AA} data. Hence, the velocity
dispersion is maybe a bit too low to fit the Ca II triplet lines
perfectly, and maybe a bit too high to fit the Ca H and K lines
perfectly.

In principle, it is possible to attribute different velocity
dispersions to different stellar populations with the modelling method
used in this paper. However, in this case the differences in velocity
dispersions cannot simply be solved this way because they are
retrieved from wavelength regions that are not far apart. 

\begin{figure}
\includegraphics[bb=50 555 250 690,scale=.8,clip=true]{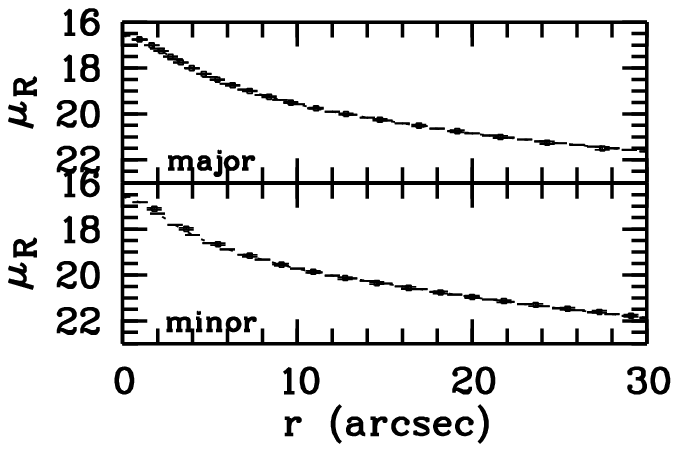}
 
\includegraphics[bb=50 95 250 730,scale=.8,clip=true]{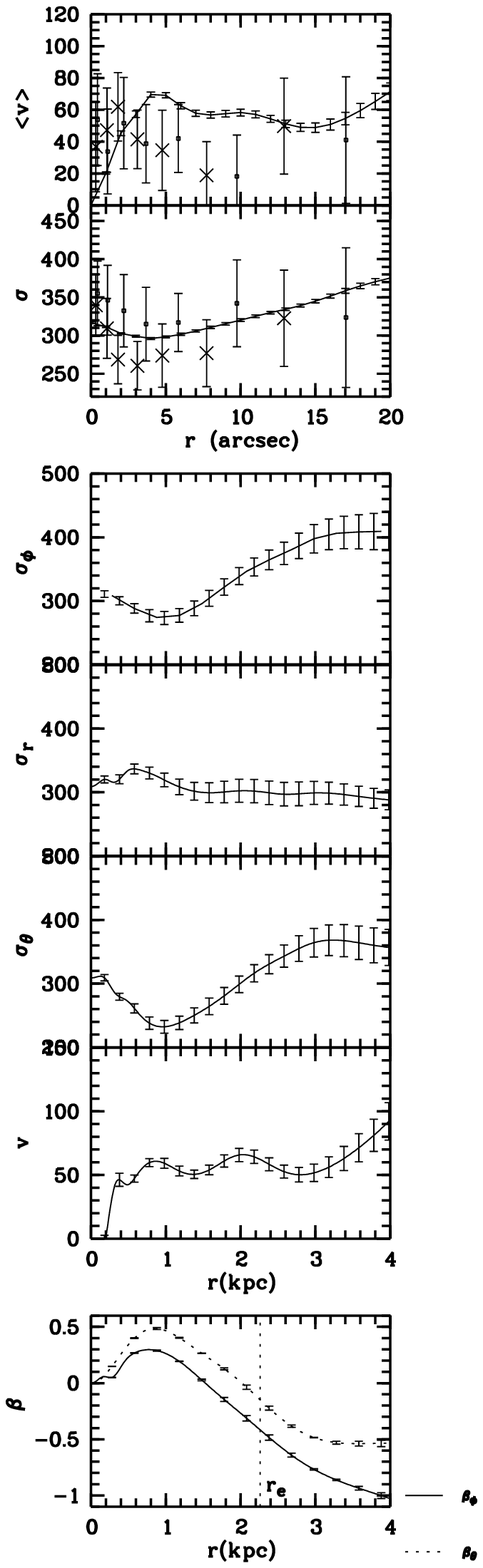} 
\caption{Some derived moments from the model based on the combined
data set. From upper to lower panel: projected density along major
axis, projected density along minor axis, projected mean velocity,
projected velocity dispersion, $\sigma_\phi$,
$\sigma_r$,$\sigma_\theta$, and the anisotropy parameters
$\beta_{\phi}$ and $\beta_{\theta}$.}  \label{modbr}
\end{figure}

Figure \ref{modbr} shows some dynamical quantities derived from the
model.  The data points that are overplotted (corrected projected mean
velocity and velocity dispersion) are taken from the near-IR data set
of the Ca II triplet (dots) and from the B data set of the Ca H and K
lines (crosses). These points are only shown for comparison, they were
not included in the fit. It is clear that the projected velocity
dispersion profile calculated from the model based on both data sets
is in general somewhat lower than the data points from the Ca II
triplet set and somewhat higher than the data points from the Ca H and
K lines in the B data set. This is again connected to the differences
in mass between the several models.

The lower panels display the spatial moments and, in the lowest panel,
the anisotropy parameters.  As can be seen from the positive values of
$\beta_\phi$ up to 1.5 kpc, the model is radially anisotropic in the
centre and becomes tangentially anisotropic further out.

\subsection{Template mixes}

\begin{figure}
\includegraphics[bb=50 128 260 220,scale=.8,clip=true]{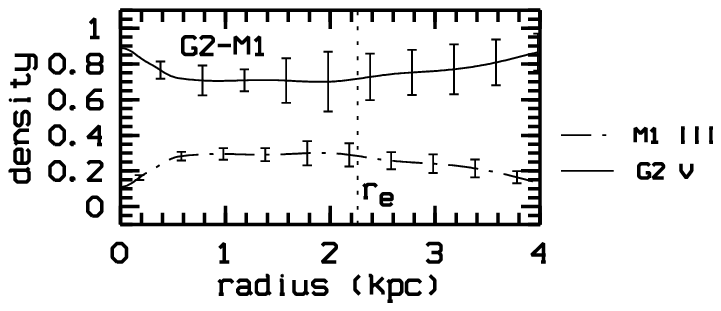} 
\caption{Template mix for the best model for the combined data set.
It is composed of G2V dwarfs and M1III giants.}  
\label{mixbrg2m1}
\end{figure}

The model composed of G2V dwarfs and M1III giants turns out to be
the best model; its template mix is shown in figure \ref{mixbrg2m1}.

\begin{figure}
\includegraphics[bb=50 65 260 220,scale=.8,clip=true]{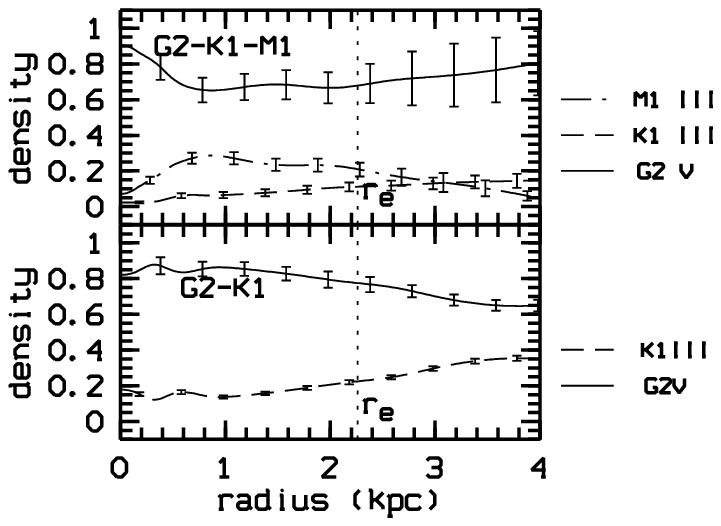} 
\caption{Template mixes containing G2V dwarfs for several acceptable
models for the combined data set.}
\label{mixbrg2}
\end{figure}

\subsubsection{Comparison with other models for the combined data set}
In fact, the model composed of G2V-K1III-M1III stars (mix presented
in upper panel of figure \ref{mixbrg2}) has the lowest value for
$\chi^2$. However, this value is only marginally lower than the
$\chi^2$ for the G2V-M1III model.  Furthermore, as can be seen from
the presentation of the mix (figure \ref{mixbrg2}), the contribution
of the K1 stars to the mix is almost negligible in the centre and is
at its maximum $( \sim 15\%)$ at the outermost radius in the figure
(there were no data available to constrain the model in this region).
The mix for the G2V-K1III-M1III model indicates that the dynamical
constraints on that population are not that strong and that the
contribution to the total DF in the inner part of the galaxy is not
that large. These are arguments in favor of presenting the G2V-K1III
model as the best fitting model.

The lower panel of figure \ref{mixbrg2} shows the stellar mix for the
model composed of G2V and K1III stars. It suggests that the
contribution of K1III stars increases outwards. This is in contrast
with the behaviour of the G2V-M1III model, where the contribution of
M1III stars is decreasing between 2 kpc and 4 kpc. In both panels of
figure \ref{mixbrg2}, G2V stars clearly contribute most to the total
density, at least 65 $\%$ for the G2V-K1III model and at least 70 $\%$
for the G2V-M1III model, with an increasing number of dwarfs toward
the centre.
\subsubsection{Comparison with models based on a single data set}

\begin{figure}
\includegraphics[bb=50 285 260 495,scale=.8,clip=true]{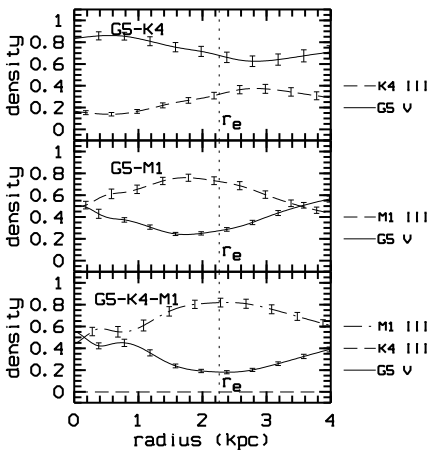} 
\caption{Template mixes containing G5V dwarfs for several acceptable
models for the combined data set.}
\label{mixbrg5}
\end{figure}

Figure \ref{mixbrg5} compares the models based on only one spectral
region with models based on the combined data set. The stellar
populations that were preferably used in the former case, are not the
ones that provide a best fit here. However, it may be instructive to
see what are the differences and the concordances. The panels that can
be compared with panels for the model based on the spectra around 8600
{\AA} in figure 15 from paper I are the ones with the G5V-K4III mix
and the G5V-K4III-M1III mix.  The G5V-K4III mix for the combined model
shows, just like for the near-IR model, a radial decrease in
contribution from G5V stars. However, the density of G5V stars is
larger for the model presented here. This is not surprising, since the
G5V stars are the dominant population in the B model, see figure
\ref{mixblue}. On the other hand, relative contributions of stellar
types in the G5V-M1III panel for the combined model look very
different from the G5V-M1III panel in figure \ref{mixblue}.  Hence, it
is no surprise that the G5V-K4III-M1III combination (but where K4III
stars are almost not used) provides not as good a model as the
G2V-K4III-M1III combination.

\subsubsection{Dynamics}
\begin{figure}
\includegraphics[bb=50 80 365 743,scale=.8,clip=true]{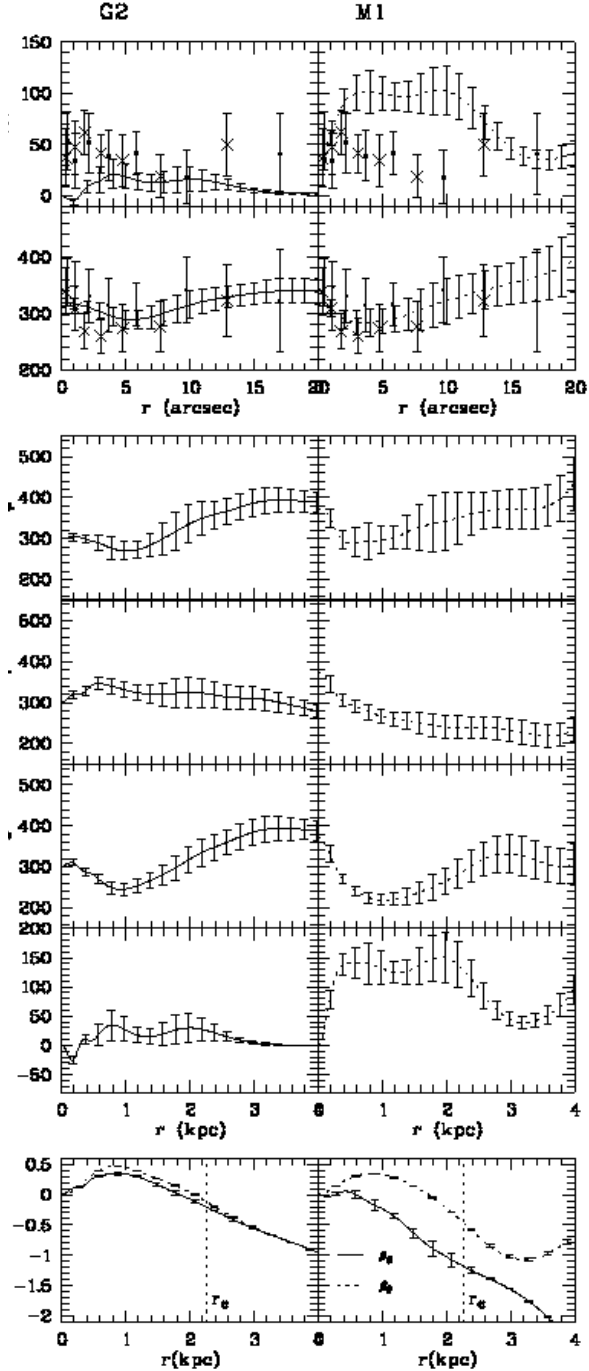} 
\caption{Dynamical quantities for separate stellar classes.
The upper panel shows the relative densities
with error bars of the populations.  The lower panels show (from
top to bottom) the projected mean velocity, the projected velocity
dispersion, $\sigma_\phi$, $\sigma_r$, $\sigma_\theta$ and the spatial
velocity for the populations (in columns).}
 \label{presbr}
\end{figure}

In figure \ref{presbr} a number of dynamical quantities are shown for
the separate populations in the best fit model for the combined data
set (composed of G2V and M1III stars). It can be compared with figure
\ref{modbr}, where the same quantities are shown for the combined DF.
The panels show (from top to bottom) the projected mean velocity, the
projected velocity dispersion, $\sigma_\phi$, $\sigma_r$,
$\sigma_\theta$, the spatial velocity and the anisotropy parameters
for the two populations (in columns).  Not surprisingly, the error
bars on the dynamical quantities are larger for the populations that
contribute less to the total density. This means that the dynamical
model is putting more constraints on the moments for the populations
that contribute more to the total density. The agreement between these
projected moments and the data retrieved with a classical technique
(overplotted in dots and crosses) is not really of great
importance. However, one would expect that for a mix of this kind
where one particular stellar type is really dominant, the dynamical
moments of this population more or less agree with the data. This is
indeed the case, as can be seen in in the first column, where the
results for the G2V stars are shown.  Form the two columns in this
figure, it becomes clear that the two populations in this model have
quite distinct kinematic properties.

\subsection{LOSVDs}
\begin{figure}
\includegraphics[bb=84 60 365 760,scale=.7,clip=true]{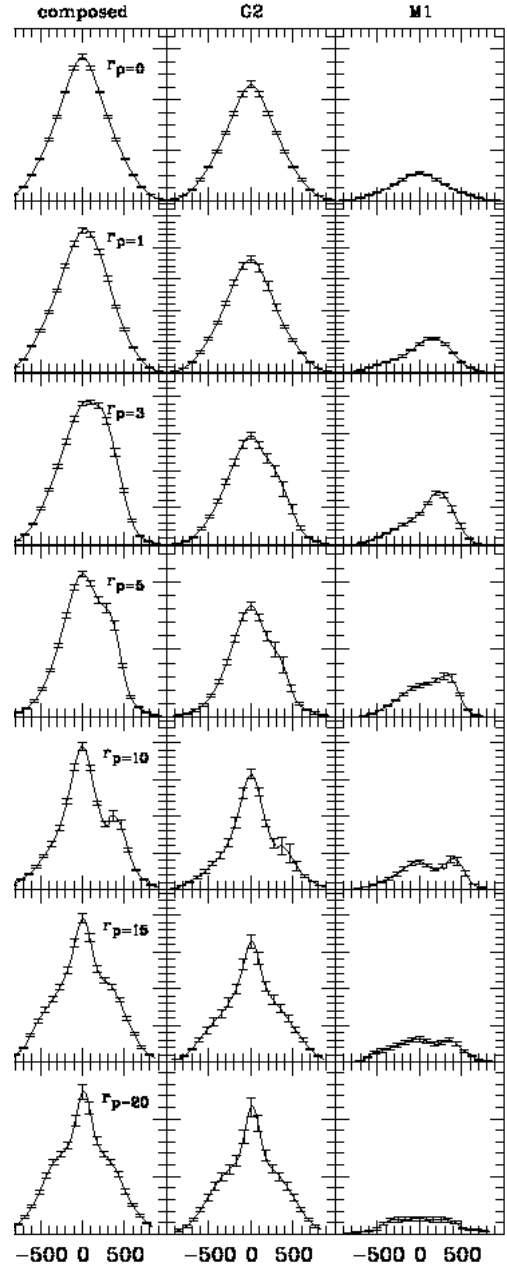} 
\caption{LOSVDs for the best fitting model based on the combined data
set. The LOSVDs are calculated on the major axis, at
0,1,3,5,10,15 and 20 arcsec from the centre (panels from top to
bottom), as indicated in the left column of the figure. From left to
right are shown: the LOSVD for the complete model, for the G2V
population, for the K1 population and for the M1III population.The
radii at which the LOSVDs are taken are indicated in arcsec the
panels. }  \label{lpbr}
\end{figure}

For the best fitting model for the combined data set, a number of line
profiles are shown in figure \ref{lpbr}. The LOSVDs are
calculated on the major axis, at 0, 1, 3, 5, 10, 15, and 20 arcsec
from the centre (panels from top to bottom), as indicated in the left
column of the figure. From left to right are shown: the LOSVD for the
complete model, for the G2V population and for the M1III population. In
general, these LOSVDs are clearly not Gaussian. The height of the
LOSVD is proportional to the contribution of the population to the
total density. 

The total LOSVD in the centre is almost symmetric, as are the LOSVDs
for G2V and M1III stars. At a projected radius of 1 arcsec, the total
LOSVD shows a slight wing toward positive velocities, whereas the
LOSVD for the G2V stars is fairly symmetric. The LOSVD for the M1III
stars is highly antisymmetric, with a strong wing toward positive
velocities. This characteristic is also reflected in the total LOSVD.
At that position (0.5 kpc), the M1III stars contribute about 30 $\%$
to the total density. Also for larger radii, it is clear that the
contribution of a small stellar class to the total LOSVD can be
important. At 10 arcsec, the bump in the total LOSVD is mainly caused
by the M1III stars, and there (at about 1.7 kpc) they also contribute
about 30 $\%$ to the total density. Also from the LOSVDs one
can see that the two stellar populations have very different
kinematics.

\subsection{Distribution functions}

\begin{figure}
\includegraphics[bb=35 210 540 800,scale=.5,clip=true]{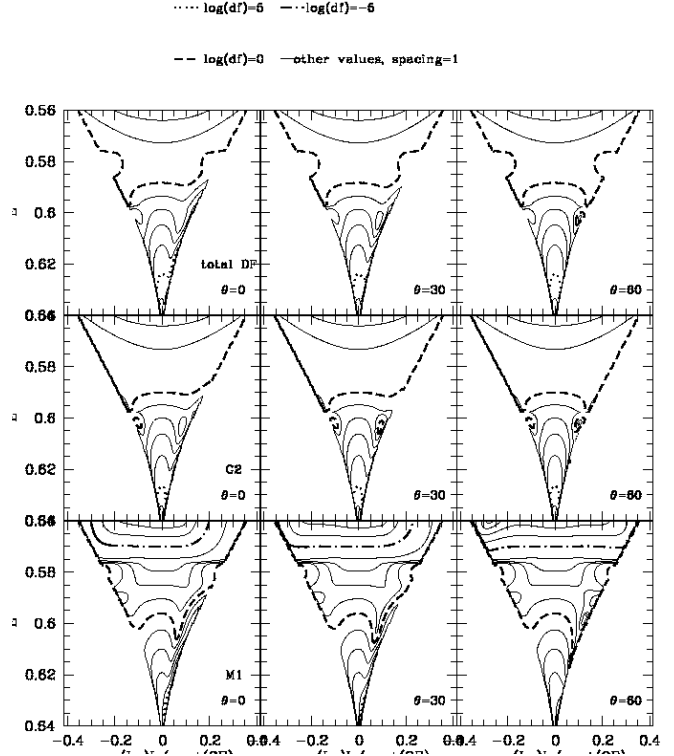}
\caption{DFs in integral space for the best fitting
model based on the combined data set.  From left to right, intersections with the equatorial plane,
with a plane tilted over $30^\circ$ and a plane tilted over $60^\circ$
with respect to the z-axis.  From top to bottom: the total DF, the DF
for the G2V class and the DF for the M1III class. }  \label{dfbr}
\end{figure}

\begin{figure}
\includegraphics[bb=35 210 540 800,scale=.5,clip=true]{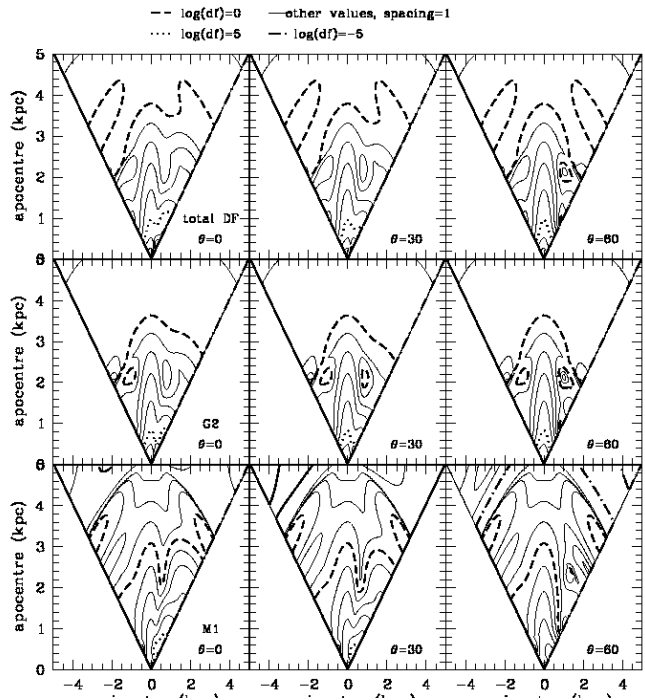}
\caption{DFs in turning point space for the best
fitting model based on the combined data set.  From left to right, intersections with the equatorial
plane, with a plane tilted over $30^\circ$ and a plane tilted over
$60^\circ$ with respect to the z-axis.  From top to bottom: the total
DF, the DF for the G2V class and the DF for the M1III class. }
\label{dfabr}
\end{figure}

Figures \ref{dfbr} and \ref{dfabr} show DFs for the best fitting
model.  Due to the fact that the model is based on a spherical
potential, the DF is built with planar orbits. The DF itself is a
function of three parameters (the energy E, the modulus of the total
angular momentum L and the vertical component of the angular momentum
$L_z$). Hence, it seems a natural choice to present this
three-dimensional function by means of sections with a plane that is
tilted with respect to the equatorial plane of the reference system
used for the modelling. In this case, the $z$-axis corresponds to
the rotation axis. All orbits in a plane tilted over an angle $\theta$
with the $z$-axis, have $L_z=L\cos\theta$. Each column in figures
\ref{dfbr} and \ref{dfabr} shows a section with a different plane: the
equatorial plane ($\theta=0^\circ$, left), a plane tilted over
$\theta=30^\circ$ (middle) and a plane tilted over $\theta=60^\circ$
(right). For a plane parallel to the z-axis, the DF is symmetric.

The plots in figure \ref{dfbr} show DFs in integral space, for the
total DF (upper row) and the DFs for the separate stellar classes. The
DF is expressed in phase space number densities. The contour plots in
figure \ref{dfabr} show DFs in turning point space, also for the total
DF (upper row) and the DFs of the separate stellar classes.  In a
sense, the representation in integral space and in turning point space
give complementary information. The representation in integral space
highlights the radial orbits, whereas the circular orbits are
squeezed. The orbits with no or very little angular momentum can be
found near the vertical axis of the panels. The opposite is true for
the representation in turning point space, see e.g. also
\citet{db}. These orbits have small differences between the values for
peri- and apocentre and can be found near the edge of the plots.

The logarithm of the DF is shown in each panel, the spacing
between the contours is 1. 

From these figures it is again clear that 
\begin{itemize}
\item For increasing tilt of the plane of section, the structure due
to the sign of $L_z$ is decreasing.
\item 
For the total DF (following the dotted contour for log(df)=5) it seems
that circular orbits are preferred, though there is also a large
fraction of radial orbits (upper left panels of figures \ref{dfbr} and
\ref{dfabr}). This is clearly due to the behaviour of the M1III stars,
that are almost all on circular orbits (lower left panels of figures
\ref{dfbr} and \ref{dfabr}). This is not surprising, since the
rotation velocity of the M1III stars is very high, where that of the G2V
stars is rather low (see figure \ref{presbr}). This preference of M1III
stars for circular orbits remains at larger radii or lower energies,
while the G2V stars gradually prefer more and more intermediate or near
radial orbits. This caused a lot of structure in the panels for the
total DF, especially in the representation in turning point space.
\item both populations have
substantially different kinematics.
\end{itemize}

\section{Discussion}

This is the first paper that reports on the application of this
modelling technique to a combination of data taken from different
wavelength regions.  Whereas the goal of simultaneously studying
kinematics and stellar populations is reached, there are still some
issues to address.

\subsection{Wavelength regions}

From paper I and section 4 in this paper it is clear that data sets
from different wavelength regions produce best models with different
potentials. This means that an accurate determination of the potential
benefits from including absorption features over a broad wavelength
range. Moreover, most of the mass estimates that are currently found
in literature are likely to be biased because of the limited
wavelength range they are based on. Combining information from
different wavelength ranges is not easy to obtain, both
observationally and on the modelling side when conventional modelling
techniques are used. Hence, we believe that the modelling method
illustrated in this paper may be a possible way to improve current
techniques.

\subsection{Technical issues}\label{techn}

The construction of a dynamical model implies that a
number of choices have to be made, that may have arisen from technical
issues originally, but some of them might raise points of discussion.

\begin{itemize}

\item How dependent is the result on the use of the specific stellar
templates?  The final model is composed of the spectra of HD21019 and
HD129902, a G2V and M1III star. It is not tested how the obtained
results change when other representatives of the G2V and M1III classes
of stars would have been used. However, the differences between
spectra of other representatives of the same stellar class are
supposed to be smaller than the differences between the spectra of
different stellar classes.  Therefore, we expect that models with
other template stars than the ones used now, will lead to the same
general conclusions, while the exact values of the DFs may be
different.  However, a similar concern may be expressed when a
standard technique is used to derive kinematic data from observed
galaxy spectra.

\item How dependent is the result on the use of the specific parts of
the spectra?  It is clear that the outcome of the population synthesis
depends on the spectral features that are used.  This is also the case
for stellar population studies that use a more common technique, like
e.g. \citet{vaz3}. These authors find in their chemo-evolutionary
population synthesis study of early-type galaxies that the metallicity
determined from the Fe lines differs from that determined from the Mg
lines.  

However, stellar population synthesis techniques are restricted to the
use of only a limited number of lines.  Classical models assuming a
single burst are not expected to explain an extended set of indices
\citep{wo} because the formation history of a real galaxy is more
complicated than that. 

For the modelling technique presented in this paper current
computational limitations (the number of data points is restricted)
put a limit on the number of absorption lines that can be
used. Moreover, the broadening of the lines due to the high velocity
dispersions in elliptical galaxies has the consequence that only the
strongest lines can be used.  For the data around 3900 {\AA}, it is
checked that the inclusion of some smaller features, at 4227 {\AA} (Ca
line) and at 4383 {\AA} (Fe line), does not change the main
conclusions of the model. Possible concerns are that the set of
template stars and dynamical components that deliver the best fit to a
small spectral region, is not adequate for fitting a broader region.

\begin{figure}
\includegraphics[bb=40 135 515 715,scale=.5,clip=]{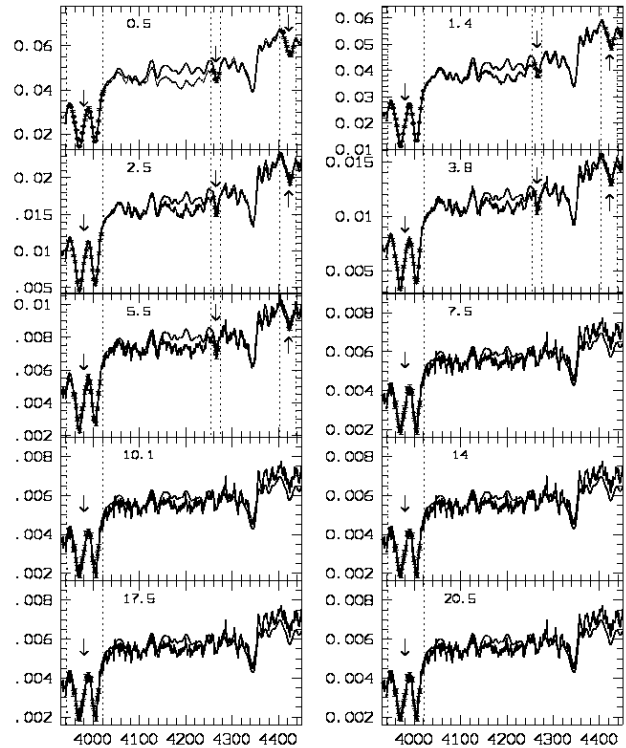}
\caption{Fit to a broader wavelength region. The
  regions included in the fit are located between the intervals
  indicated by dashed lines and are indicated with an arrow. The
  projected radii of the positions where the spectra are taken are
  indicated in the panels. }
\label{verder}
\end{figure}

The fit to this extended B data set is shown in figure \ref{verder},
the spectral lines of interest are between vertical dotted lines. The
fit to the Ca 4227 and Fe 4383 lines (first five panels) is within the
error bars. For larger radii, the lines were not included in the fit,
but even there, the Ca 4227 line is also good represented by the
model spectra. For the Fe 4383 line, the difference between the
original and model spectrum is larger. The main reason for this is
probably the fact that during the fit, the normalization of the
spectra is only applied for the parts that are included in the fit.

This shows that in this case the extended data set can be modelled
with the stellar templates and dynamical components that give a good
fit for the smaller data set. Hence, we expect that for a model based
on the strongest lines (the ones that are used now) and some weaker
lines the results on the stellar mix and the dynamical state of the
galaxy and populations are robust, though the values of the DFs and
the derived moments may be slightly different.

\item How realistic are the error bars on the model and the derived
quantities?  The error bars on the model depend on the errors on the
data and on the components that are included in the model.  The error
bars on the models are derived from the Hessian matrix involved in the
quadratic programming. These error bars have a statistical meaning, if
the noise on the spectra is Poisson noise and if the data points are
weighted by their errors. But after standard data reduction, rebinning
and sky subtraction, the noise on the spectra is no longer pure
Poisson noise, neither are the data points independent (see
\citet{db2}). However, the use of a quadratic programming assumes
independent data points.

If a component goes to zero, also the error on that component goes to
zero. This explains why e.g in figure \ref{presbr}, for the G2V stars,
the error on the velocity is zero for larger radii.  It is also
possible that the error on the dispersion is lower in outer regions.
This happens when less components contribute in the outer regions than
for the inner regions.  There will be larger constraints on these
fewer components, hence the error bars will be smaller.  This
indicates that the error bars calculated from the model will
underestimate the real errors.

\end{itemize}

\subsection{Formation history}

The photometry (see paper I) revealed a dust disk around the centre of
NGC~3258. Unfortunately, our B spectra do not extend far enough to
check the presence of any [OII] emission. Central dust disks are found
in a number of galaxies. Most of these galaxies are powerful radio
sources (e.g. NGC~7052 \citet{vdm2}, NGC~4261 \citet{jon}) or show
clear signs of kinematically distinct cores \citep{car}. This is not
the case for NGC~3258.  In X-rays, there is no sign of major activity
in the centre of this galaxy, the measured flux is consistent with the
$L_x-\sigma$ relation pointed out by \citet{mah}.  Hence, it is not
very probable that the centre harbours an (at the moment) active
nucleus.

Due to the limited spatial resolution of our data, we detect no
kinematic signature of a central massive black hole, although
following the scaling relation of \citet{ferm}, a central black hole
of about $10^9 M_\odot$ could be present in the galaxy.


Moreover, the disk seems to be aligned with the major axis of the
galaxy, which tends to be the case in more galaxies \citep{tran}.
This may well point at an internal origin of the dust, or of a capture
long enough ago so that no obvious signs of interaction are left.

The B-R colour indicates a reddening toward the centre and the
population synthesis points suggests that the number density of dwarfs
is increasing from 0.5 kpc inwards. Hence this reddening may be caused
by a combination of dust effects and higher metallicity, consistent
with findings of e.g. \citet{ka}. Also the Mg2 indices \citep{gol}
point toward higher metallicity in the centre.  This also suggests
that the star formation in the centre went on longer than in the outer
parts of the galaxy.  This does fit in a formation scenario as
outlined and simulated by \citet{mori}, where the shrinking of a
protogalaxy increases the density in the centre so that active star
formation ignites. In the centre of the galaxy, cycles of newborn and
dying stars continue and cause the metallicity to increase. In the
outer regions there is not enough gas to continue star formation in an
enriched environment.

The central dust disk reported in paper I probably harbours the hot
dust component of $48 M_\odot$ \citep{fer}.  This could be interpreted
as a star formation region or as the signature of a black hole. To
solve this issue of the hot dust, further observations would be
required.


It seems possible that NGC~3258 is observed in a stage in the cycle
where there is no (or very low) star formation. The increase of dwarfs
toward the centre may be a remnant of a past star formation period.
The fact that there is still dust in the centre and still some gas,
conform the weak [OII] and HI emission, leaves the possibility open
that this galaxy will go through another star forming episode in the
future. If there is indeed a black hole in the centre, following a
suggestion of \cite{tran}, the present period of inactivity may be due
to an at the moment inefficient fuel accretion.

The dynamical analysis of the two stellar populations in the model
based on the Ca H and K lines and the Ca II triplet lines indicates that
the old population of M1III giants contains a lot of rotation, while the
younger G2V dwarfs show little or no rotation. 
The difference in rotation between the two populations suggests that
the bulk formation of dwarfs occurred at a moment when the material
they are made of had less angular momentum.  Also \citet{dok} suggest
that dust should have some way of losing angular momentum. A possible
way to do this might be through dissipationless merging. \cite{zurek}
found in their simulations that dissipationless merging forces the
cores of the merging galaxies into the core of the final product, in
the process of which cores become more bound and lose angular
momentum.

\section{Conclusions}
In this paper we present a method to analyse spectra from elliptical
galaxies in order to retrieve dynamical information and a stellar
population synthesis at the same time. The method originates from the
field of dynamical modelling and is outlined by \citet{dr}. The main
idea is that the different stellar populations that contribute to the
integrated galaxy spectrum do not necessarily share the same kinematic
characteristics. Hence, in terms of dynamical modelling, they should
be given different distribution functions. Hence, our aim is to
find a spatial distribution of the different stellar populations in
the galaxy under study, while the dynamical constraints are obeyed. 

Until very recently, stellar template spectra observed with the
same instrumental setup as the galaxy spectrum were indispensable to
extract kinematical information from the galaxy spectra. However,
facing the realities concerning the obtainment of observing time on
heavily oversubscribed observatories, it is simply impossible to
observe a large number of template stars, covering a wide range of
spectral types, ages, and metallicities. Nor do the galaxy spectra,
usually covering only a limited spectral region with only a handful of
strong absorption lines, allow to constrain the dynamical states of a
large number of distinct stellar populations. Since the template
spectra need to have the same instrumental characteristics as the
galaxy spectra, it is very difficult to extend an observed set of
templates with archival data. 

If stellar templates are used, the method presented in this paper does
not explicitly take an evolutionary status into account and is akin to
the empirical spectral synthesis method pioneered by Pickles
(\citet{pivi}). We do conceed that the present method can only show
its full potential if realistic template spectra are used. With the
work of Vazdekis (\citet{vaz2}), SSP spectra have become available
that potentially can replace stellar templates. These SSP spectra have
high spectral resolution and can be transformed to the instrumental
characteristics of the observed galaxy spectra (see Falc\'on-Barroso
(\citet{fal}) for an first application). However, they still have to
prove their usefulness in the field of galaxy dynamics and much more
applications to real data will have to be done before they can
completely replace observed stellar templates. Although this holds
great promises for the future, stellar templates will be in use for
some time to come. Our modeling method makes the best possible use of
the kinematical and stellar-mix information in the spectra and nothing
prevents us from replacing observed templates by SSP spectra, opening
the door to a complete assessment of the dynamical and chemical state
of a stellar system.

Such models can be used to obtain more information than immediately
available from the observations. An other important advantage is that
dynamical models are suitable for comparisons between individual
galaxies (e.g. \cite{db}), whereas evolutionary models are less
appropriate for this \citep{wo}. However, the latter have the
advantage that they consider an evolution in time. With the method
presented here it is possible to combine spectral features coming from
different wavelength regions and even different observational
setups. This is an important advantage in comparison with modelling
methods that use LOSVDs or kinematic parameters that are derived from
the spectra.

More specific conclusions are :
\begin{itemize}
\item Potential.

The potential derived from the Ca II triplet or the Ca H and  K lines is
different. The slope of the dark matter halo is the same, but the
total amount of mass is different. This conclusion is based on the
modelling of the spectra, but the modelling of the kinematic profiles
would probably also lead to different potentials. The total amount of
mass inferred from the model based on the combined data sets lies in between the
masses of the models based on only one spectral region. These three
models give an estimate for the total mass at 2 $r_e$ (4.5 kpc) of
$4.5\times10^{11} M_\odot\pm 0.5 \times10^{11}M_\odot$ (lower limit
from B model, upper limit from near-IR model). At this radius 37 $\%$
of the mass is in luminous matter.

This indicates that establishing a potential for a galaxy may be
dependent on the wavelength range used for the modelling. Here,
the total mass included in the B and near-IR models at 2 $r_e$
differs as much as $10^{11} M_\odot$ (about 20 to 25 \% of the total
mass) depending on the wavelength range that is used. If
multiwavelength information is indeed required to determine the
potential of elliptical galaxies, this modelling method has
non-negligible advantages.  It is clear that in such a case only
modelling methods can be used where it makes sense to combine
information from different wavelength regions.

The modelling did not require to include a black hole in the centre.

\item Stellar populations.

The stellar population mixes from the models based on the Ca II
triplet, the Ca H and K lines and the combined data sets are different.
For the models based on one spectral region, this can be expected,
since the sensitivity of the spectral features that are used is
different.

However, the stars that give the best results for these models
(G5V-K4III for the Ca II triplet and G5V-M1III for the Ca H and K
lines) do not provide the stellar mix that gives the best results for
the model based on both spectral regions. The best model is in this
case composed of G2V and M1III stars, more than 75 $\%$ of the
stars are dwarfs.

Toward the centre there is an increase in dwarfs, this result is also
seen in the B model and is consistent with results obtained by
\citet{trag2}. This indicates that the reddening toward the centre is
caused by a metallicity gradient. Part of the reddening may be caused
by the dust disk (see paper I).
\item Internal dynamics 

All three models are isotropic in the very centre, radially anisotropic in
an intermediate region (till roughly $1 r_e$ for the near-IR model and
roughly $0.5 r_e$ for the B model and till roughly $0.66 r_e$ for the
combined model) and are tangential anisotropic at large radii.

The model for the combined data set is studied in more detail and
clearly attributes substantially different kinematic behaviour to the
two classes of stars used for the combined spectra. In this case,
almost all rotation is delivered by the giants.

\end{itemize}

\section{Appendix}

\section*{Acknowledgments}
VDB acknowledges financial support from FWO-Vlaanderen.  WWZ
acknowledges the support of the Austrian Science Fund (project P14783)
and the support of the Bundesministerium f\"ur Bildung, Wissenschaft
und Kultur.

\bsp

\label{lastpage}

\end{document}